\RequirePackage{fix-cm}
\documentclass[twocolumn,epjc3]{svjour3}
\smartqed  
\RequirePackage{graphicx}
\journalname{Eur. Phys. J. C}
\begin{document}

\title{ Status of quarkonia - like negative and positive parity states in a relativistic confinement scheme
}


\author{Tanvi Bhavsar\thanksref{e1,addr1}
        Manan Shah \thanksref{e2,addr2}
        \and
P C Vinodkumar\thanksref{e3,addr1}
}

\thankstext{e1}{e-mail: tanvibhavsar1992@yahoo.com}
\thankstext{e2}{e-mail: {mnshah09@gmail.com}}
\thankstext{e3}{e-mail: {p.c.vinodkumar@gmail.com}}

\institute{Department of Physics, Sardar Patel University, Vallabh Vidyanagar, 388120, India \label{addr1}
           \and
            P. D. Patel Institute of Applied Sciences, CHARUSAT, Changa, 388421, India \label{addr2}
}

\date{Received: date / Accepted: date}

\maketitle

\begin{abstract}
Properties of quarkonia - like states in the charm and bottom sector have been studied in the frame work of relativistic Dirac formalism with a linear confinement potential. We have computed the mass spectroscopy and decay properties (vector decay constant and leptonic decay width) of several quarkonia - like states. Present study is also intended to identify some of the unexplained states as mixed P-wave and mixed S-D wave states of Charmonia and Bottomonia. The results indicate that the X(4140) state  can be an admixture of two P states of charmonium. And the charmonium like states X(4630) and X(4660) are the admixed state of S - D waves. Similarly, the $X(10610)$ state recently reported by Belle II can be a mixed P - states of bottomonium.
 In the relativistic framework we have computed vector decay constant and the leptonic decay width for S wave charmonium and bottomonium. The leptonic decay width for the $J^{PC} =  1^{--}$  mixed states are also predicted. Further, both the masses and the leptonic decay width are considered for the identification of the quarkonia-like states.
\keywords{Heavy quarkonia \and Relativistic quark model \and  Mesons}
 \PACS{14.40.Pq \and 12.39.Ki \and 14.40.-n}
\end{abstract}

\section{Introduction}
\label{intro}
In recent years remarkable experimental progress has been achieved in
the investigation of charmonium-like and
bottomonium-like states. The latest experimental results on heavy flavour hadrons have gained renewed interest in heavy flavor Physics \cite{N. Brambilla,pdg2017} to understand the properties of strongly interacting hadrons. Conditions seemed to be very different for spectra above and below
flavor threshold region. In the region above the open charm threshold, number of
charmonium-like states (the so-called "X Y Z" states) have been discovered with unusual properties. These states might be exotic states, mesonic molecules or multi quark states
\cite{N. Brambilla}.

\begin{table*}
\begin{center}
\tabcolsep5pt
   \small
\caption{Experimental status of some of the negative parity and positive parity quarkonia - like states.} \label{tab1}
\begin{tabular}{c c c c c}
\hline
Exp. State & Exp.mass (MeV) & $J^P$ & Process (mode) & Experiment \\
\hline

Y (4008) &$4008^{+121}_{-49}$ &  $1^-$ & $e^+e^-\rightarrow \gamma(\pi^+\pi^-J/\psi)$ & Belle \cite{cz} \\

$\psi(4160)$ & $4191 \pm 5$ &   $1^-$ & $e^+e^-\rightarrow \eta J/\psi$ & Belle \cite{xlw}\\

Y (4220) &$4222.0\pm3.1\pm1.4$ &  $1^-$ & $e^+e^-\rightarrow \gamma(\pi^+\pi^-J/\psi)$ & BESIII \cite{bes} \\

Y (4260) & $4263^{+8}_{-9}$ &  $1^-$ & $e^+e^-\rightarrow \gamma(\pi^+\pi^-J/\psi)$& BABAR\cite{ba1,ba2}CLEO\cite{qne} , Belle \cite{cz} \\

&&&$e^+e^-\rightarrow (\pi^+\pi^-J/\psi)$ & CLEO \cite{te}\\

&&&$e^+e^-\rightarrow (\pi^0\pi^0 J/\psi)$& CLEO\cite{te}\\

Y (4330) &$4320.0\pm10.4\pm7.0$ &  $1^-$ & $e^+e^-\rightarrow \gamma(\pi^+\pi^-J/\psi)$ & BESIII \cite{bes} \\

Y (4360) & $4361 \pm 13$ & $1^-$ & $e^+e^-\rightarrow \gamma(\pi^+\pi^-\psi(2S))$ & BABAR \cite{ba3} , Belle \cite{xl} \\

X(4630) & $4634^{+9}_{-11}$
& $1^-$ &$e^+e^-\rightarrow \gamma(\wedge^+_c\wedge^-_c)$& Belle \cite{gp1}\\

Y (4660) &4664$\pm$12 & $1^-$&$e^+e^-\rightarrow \gamma(\pi^+\pi^-\psi(2S))$& Belle \cite{xl}\\

$Y_b(10888)$ &10888.4$\pm$3.0 & $1^-$& $e^+e^-\rightarrow \gamma(\pi^+\pi^-\Upsilon(nS))$& Belle \cite{kfc1,kfc2}\\

$X(10610)$ &10609 $\pm$ 4.0 & $1^+$ &$ e^+ e^- \rightarrow \Upsilon(2S)/\Upsilon(3S)\pi^0\pi^0 $ & Belle \cite{pdg2017} \\

$h_c(1P)$ & 3525.41 $\pm$ 0.16 & $1^+$ & $\psi(2S) \rightarrow \pi^0 (\gamma \eta_c(1S))$ & CLEO \cite{JLR,S} \\

$h_b(2P)$ &$10259.8^{+1.5}_{-1.2}$
& $1^+$ & $\Upsilon (5S) \rightarrow \pi^+\pi^- (...) $ & Belle \cite{IA}\\

$X(3940)$ &$3942^{+7}_{-6}\pm 6$
& $?^?$ & $e^+ e^- \rightarrow J/\psi X $ & Belle \cite{pdg2017}\\

$X(4020)$ &$4025.5^{+2.0}_{-4.7}\pm 3.1$
& $?^?$ & $e^+ e^- \rightarrow (D^\ast \bar{D}\ast)^0 \pi^0 $ & BESIII \cite{pdg2017}\\

$X(4140)$ &$4143\pm {+2.9}\pm {1.2}$
& $?^?$ & $B^+ \rightarrow J/\psi \phi K^+ $ & CDF \cite{pdg2017}\\

$X(4350)$ &$4350.6^{+4.6}_{-5.1}\pm 0.7$
& $?^?$ & $e^+ e^- \rightarrow e^+ e^- J/\psi \phi $ & BELL \cite{pdg2017}\\
\hline
\end{tabular}
\end{center}
\end{table*}

 Most of these unknown states do not fit in the standard
charmonium and bottomonium spectra \cite{skc,HAI-BO LI}.
All the narrow charmonium states below the open-charm threshold have been observed
experimentally and their mass spectrum can be well described by potential models \cite{C. Patrignani}. We have sufficient knowledge of $\eta_c(1S)$ and $\eta_c(2S)$.
The BESIII/BEPCII facility in Beijing, has shed more light on these spin-singlet states by collecting a new record of $\psi(3686)$ decays
in electron-positron annihilations \cite{Zahra Haddadi}. Recently, BESIII showed that the Y(4260) is split up in to two resonant states: one with a mass of $4222.0 \pm 3.1 \pm 1.4 $  $MeV/c^2$ and the other with a
mass of $4320.0 \pm 10.4 \pm 7.0 $ $MeV/c^2$ in their cross section measurement
of $e^+e^− \rightarrow π^+π^− J/\psi $ for center of mass energies from $\sqrt{s}$ =
3.77 to 4.60 GeV \cite{bes}. Large amount of data on charmonium and bottomonium production is available at RHIC \cite{A. Adare,ca,la1,la2,la3} and at the LHC
 \cite{B. Abelev,S. Chatrchyan,1S. Chatrchyan,2S. Chatrchyan,3S. Chatrchyan,E. Abbas,B.B. Abelev} significantly extending our understanding of quarkonium production in
deconfined matter \cite{aan}. To understand all these, we have to go beyond the conventional quark or quark anti-quark bound systems. There are various issues related to higher excited states which are still to be resolved. In this context, phenomenological models either non-relativistic quark model (NRQM) or the relativistic quark model have been developed to study the properties of heavy mesons (Charmonium and Bottomonium) \cite{pcv2005,sjp,de}.\\

In the present study we compute the masses of charmonium -like and bottomonium-like states in a relativistic frame work. The mass spectroscopy
of charmonium and bottomonium states are observed  experimentally with
 high accuracy \cite{pdg2017}. But the  masses of S-wave charmonium states beyond 3S and
the bottomonium states beyond 4S are not very well resolved. There are many other X,Y and Z states above the $c\bar{c}$ and $b\bar{b}$ threshold which also require to be identified.
For example  $\psi (3770)$, $Y(4008)$, $Y(4220)$, $Y(4260)$, $Y(4330)$, $Y(4360)$, $X(4630)$, $Y(4660)$, $X(10610)$, $Y_b (10880)$ etc. have the same  $J^{PC}$
value $1^{--}$ and justify to be one of the quarkonia-like states \cite{S. Eidelman}.
According to PDG 2016, the earlier states have been now renamed: Y(4260) as X(4260), Y(4360) as X(4360), Y(4660) as X(4660) and $Y_b(10888)$ as $\Upsilon(10860)$. Some of these states can be either hidden charm (X, Y, $Z_c$) or hidden bottom ($Y_b$ and $Z_b$) states
and are located above the open charm or open bottom threshold. It is known that their decay properties can also throw light on their identity. Thus we incorporate to compute leptonic decay properties of these $1^{- -}$ states for comprehensive understanding of these quarkonia-like states. The ultimate goal of this study is to describe the status and properties of the X, Y, Z states with the help of phenomenological model. However, this task is quite challenging as more and more new quarkonia-like states are observed.

\subsection{$J^P$ = $1^-$ States}

In ISR (Initial-State Radiation) process, BaBar observed peaks near 4300 $MeV/c^2$ in the $\pi^+\pi^-J/\psi$ and $\pi^+\pi^-\psi'$ channels. The partial widths for these two decay channels are  larger than that required to observe charmonium states. As these states are produced via the ISR process, they have $J^P$ = $1^-$ \cite{Stephen}. Some of these $1^-$ states are listed in Table \ref{tab1}.

\subsection{$J^P$ = $1^+$ States}

\paragraph*{} X(3872) was observed by BELLE \cite{pdg2017}, then after it was
confirmed by BABAR \cite{12} and its $J^P$ value $1^+$ was determined by LHCb \cite{lhcb}.
Other unknown state Z(4475) which was produced in a charmonium-rich B
meson weak decay process, has a mass near to the  excited
charmonium. It is believed that this state is
a strong candidate for hidden charm tetraquark state \cite{z}. From experimental observations it might be possible that a state with $J^P$ value $1^+$ can be a molecular or tetra quark kind of state. In the present study we also look at these states as admixture of P-waves of quarkonium states. Some of these $1^+$ states are also listed in Table \ref{tab1}. The present study based on relativistic Dirac formalism is an attampt to understand stand the quarkonia like states below and above the $c\bar{c}$ and $b\bar{b}$ states.\\

 The paper is organized as follows. In Sec.II we briefly discuss our relativistic quark model  based on the Dirac formalism.  In Sec.III, the leptonic decay width and decay constant of $1^{--}$ quarkonia are computed and the results are compared with the available experimental results and with other theoretical model predictions. Sec.IV contains mixing of two nearby mesonic states and predicted the status of experimentally known unresolved negative parity and positive parity states. The summary and conclusion of the present study is presented in Sec.V.\\

\section{Theoretical frame work}


One of the most successful way to construct the quarkonium system is to solve  Dirac equation for the quark and anti quark in a confinement potential. For the present study we have considered the confinement through a linear potential.
The form of the model potential is expressed as,

\begin{equation}\label{eq:a}
V(r)= \frac{1}{2} (1+\gamma_0) (\lambda r^{1.0}+V_0)
\end{equation}

Where,
$\lambda$ is the strength of the confinement part of the potential \cite{pcv}. $V_0$ is a constant  negative potential depth \cite{epj,prd2014,prd2016}.\\

The wave function which satisfy Dirac equation with a general potential is given by \cite{aruldhas,greiner},

\begin{equation}\label{eq:b}
(\vec{\alpha}\cdot\vec{p} + m_Q) \psi_q (\vec{r}) = \left[ E_q - V (r) \gamma_0 \right] \psi_q (\vec{r}),
\end{equation}

\begin{equation}\label{eq:c}
[\gamma^0 E_q - \vec{\alpha}. \vec{p} - m_q - V (r)]\psi_q (\vec{r}) = 0,
\end{equation}

where
\begin{equation}\label{eq:d}
\alpha = \left(
    \begin{array}{cc}
     0      & \sigma \\
     \sigma & 0
    \end{array}
  \right); \qquad
\\  \gamma^0 =  \left(
    \begin{array}{cc}
     1      & 0\\
     0      & -1
    \end{array}
  \right); \qquad
\\ \gamma^i = \left(
    \begin{array}{cc}
     0      & \sigma_i \\
     -\sigma_i & 0
    \end{array}
  \right)
\end{equation}

V(r) is a potential which consist scalar + vector part.
Main feature to use scalar plus vector potential is that it is
applicable for the bound states of both mesons and baryons \cite{pcv}.\\

\begin{table*}
\begin{center}
\tabcolsep 5pt
\caption{Model Parameters fitted in our model for the Charmonium and bottomonium systems.} \label{tab2}
\begin{tabular}{c c c}
\hline
System Parameters     & $b\bar{b}$        &       $c\bar{c}$  \\
\hline
Quark mass (in GeV/$c^2$)      & $m_{b/\bar{b}} $ = 4.67& $m_{c/\bar{c}}$ = 1.27 \\
$V_0$(GeV)&   $\frac{-0.246}{(n + 1)^{1.42}}$ & $\frac{ -0.146}{(n + 1)^{3.01}}$\\
Potential strength ($\lambda$) $(GeV^2)$& $0.18$ & $0.084$\\
\hline
\end{tabular}
\end{center}
\end{table*}

The solution of Dirac equation can be written as two component (positive and negative energies in the zeroth order) form as \cite{greiner,epj,prd2014,prd2016},\\
\begin{equation}\label{eq:d}
\psi_{nlj}(r) = \left(
    \begin{array}{c}
      \psi_{nlj}^{(+)} \\
      \psi_{nlj}^{(-)}
    \end{array}
  \right)
\end{equation}
where
\begin{equation}\label{eq:e}
\psi_{nlj}^{(+)}(\vec{r}) = N_{nlj} \left(
    \begin{array}{c}
      i g(r)/r \\
      (\sigma.\hat{r}) f(r)/r
    \end{array}
  \right) {\cal{Y}}_{ljm}(\hat{r})
\end{equation}
\begin{equation}\label{eq:f}
\psi_{nlj}^{(-)}(\vec{r}) = N_{nlj} \left(
    \begin{array}{c}
      i (\sigma.\hat{r}) f(r)/r \\
        g(r)/r
    \end{array}
  \right) (-1)^{j+m_j-l} {\cal{Y}}_{ljm}(\hat{r})
\end{equation}
and $N_{nlj}$ is the overall normalization constant \cite{greiner,epj,prd2014,prd2016}. The normalized
 spin angular part is expressed as
\begin{equation}\label{eq:g}
{\cal{Y}}_{ljm}(\hat{r}) = \sum_{m_l, m_s}\langle l, m_l, \frac{1}{2}, m_s| j, m_j \rangle Y^{m_l}_l \chi^{m_s}_{\frac{1}{2}}
\end{equation}
Here the spinor $\chi_{\frac{1}{2}{m_s}}$ are eigenfunctions of the spin operators \cite{greiner,epj,prd2014,prd2016},
\begin{equation}\label{eq:h}
\chi_{\frac{1}{2} \frac{1}{2}} = \left(
    \begin{array}{c}
      1 \\
      0
    \end{array}
  \right) \ \ \ , \ \ \ \ \chi_{\frac{1}{2} -\frac{1}{2}} = \left(
    \begin{array}{c}
      0 \\
      1
    \end{array}
  \right)
\end{equation}
The reduced radial part $g(r)$ and $f(r)$ of the Dirac spinor $\psi_{nlj}(r)$ are the solutions of the equations given by \cite{greiner,epj,prd2014,prd2016},\\
\begin{equation}\label{eq:i}
\frac{d^2 g(r)}{dr^2}+\left[(E_{D}+ m_q)[E_{D} - m_q - V(r)]-\frac{\kappa(\kappa + 1)}{r^2}\right] g(r) = 0
\end{equation}
and
\begin{equation}\label{eq:j}
\frac{d^2 f(r)}{dr^2}+\left[(E_{D}+ m_q)[E_{D}- m_q - V(r)]-\frac{\kappa(\kappa-1)}{r^2}\right] f(r) = 0
\end{equation}
it is appropreate to define a new quantum number $\kappa$ \cite{greiner,epj,prd2014,prd2016} as,
\begin{equation}\label{eq:k}
  \kappa = \left \{ \begin{array}{c}
      -(l+1) = -(j+\frac{1}{2})      \ \ \  for \ \ \   j = l + \frac{1}{2}\\\\

      l = (j+\frac{1}{2})          \ \ \   for \ \ \  j = l - \frac{1}{2}
    \end{array}\right.
\end{equation}

On converting these equation into dimensionless form \cite{aruldhas,epj,prd2014,prd2016} as,
\begin{equation}\label{eq:l}
\frac{d^2 g(\rho)}{d\rho^2}+\left[\epsilon - \rho^{1.0} - \frac{\kappa (\kappa + 1)}{\rho^2}\right] f(\rho) = 0
\end{equation}

\begin{equation}\label{eq:m}
\frac{d^2 f(\rho)}{d\rho^2}+\left[ \epsilon - \rho^{1.0} - \frac{\kappa (\kappa - 1)}{\rho^2}\right] g(\rho) = 0
\end{equation}

where $\rho= \frac{r}{r_0}$ is a dimensionless variable with suitably chosen scale  factor
 $r_0 =\frac{r}{[(E+m)\lambda] ^\frac{-1}{3}}$ and corresponding  energy eigen value is given by \cite{epj,prd2014,prd2016},

\begin{equation}\label{eq:n}
\epsilon = (E_D-m_q-V_0)(m_q+E_D)^\frac{1}{3} \lambda^\frac{-2}{3}
\end{equation}

The solution of $f(\rho)$ and $g(\rho)$ are normalized to get \cite{epj,prd2014,prd2016},

\begin{equation}\label{eq:o}
\int^\infty_0 \left[f^2(\rho)+g^2(\rho)\right]d\rho = 1
\end{equation}

Now the wave function for quarkonium system can be constructed by using positive and negative energy
solutions of Dirac equation. Mass of particular Quark-Anti quark system can be written as \cite{epj,prd2014,prd2016},

\begin{table*}
\begin{center}
\tabcolsep 8pt
   \small
\caption{S-wave mass spectrum for $b\bar{b}$ and $c\bar{c}$ bound states(in MeV).} \label{tab3}
\begin{tabular}{c c c c c c c c}
\hline
&&&Bottomonium\\
\hline

nL & State &Present&  Experimental \cite{pdg2017}& \cite{prd2012}& \cite{radford} & \cite{Wei-Jun Deng} & \cite{aeb}\\
\hline

1S   &$1{^3S_1}$& 9460.99 & 9460.30 $\pm$ 0.26  &9460.43 & 9460.38 &  9460&  9608  \\
    &$1{^1S_0}$& 9390.7 &  9399.0 $\pm$ 2.3  & 9392.38& 9392.91&9390&  9607   \\

2S   &$2{^3S_1}$&  10024.1& 10023.26 $\pm$ 0.31 & 10023.80 & 10023.3&10015&  10023.3 \\
     &$2{^1S_0}$&  9999.3 &  $\ldots$    & 9990.88 & 9987.42&9990&   $\ldots$    \\

3S  &$3{^3S_1}$ & 10356.2 & 10355.2  $\pm$  0.5 & 10345.80& 10364.2&10343& 10353.3 \\
     &$3{^1S_0}$& 10325.3 &  $\ldots$  &  10323.40 & 10333.9&10326&  $\ldots$    \\

4S   &$4{^3S_1}$& 10576.2 &   10579.4 $\pm$  1.2 & 10575.20 & 10636.4&10597& 10580   \\
     &$4{^1S_0}$&  10554.4 &    $\ldots$    & 10558.30 &  10609.4 & 10584& $\ldots$   \\

5S   &$5{^3S_1}$& 10758.5 &  $\ldots$ &10755.40 & $\ldots$ &10811& 10865 \\
     &$5{^1S_0}$&  10738.4 &    $\ldots$     & 10741.40 & $\ldots$ &10800& $\ldots$    \\
\hline

&&&Charmonium\\
\hline

nL & State & Present  &  Experimental \cite{pdg2017} & \cite{prd2012}& \cite{Bai-Quing Li} & \cite{nr}  & \cite{aeb} \\
\hline

1S   &$1{^3S_1}$&  3096.7 & 3096.90 $\pm$ 0.006& 3097.14& 3097 & 3090&  3096.9  \\
    &$1{^1S_0}$ & 2977.8 &2983.4 $\pm$ 0.5 & 2979&2982&    2979     \\

2S   &$2{^3S_1}$& 3684.4 &3686.097 $\pm$ 0.025  & 3689.95&  3673& 3672& 3686   \\
     &$2{^1S_0}$& 3630.5 & 3639.2 $\pm$ 1.2 & 3633.49& 3623 & 3630 &  $\ldots$     \\

3S  &$3{^3S_1}$ & 4022.4 & $\ldots$ &4030.32 &4022 & 4072&  3769.9   \\
     &$3{^1S_0}$& 3990.8 & $\ldots$ & 3991.99 &3991& 4043&    $\ldots$     \\

4S   &$4{^3S_1}$& 4266.4 & $\ldots$  & 4273.49 & 4273& 4406&   4040  \\
     &$4{^1S_0}$& 4262.1& $\ldots$ & 4244.11& 4250 & 4384&  $\ldots$      \\

5S   &$5{^3S_1}$&  4441.5&  $\ldots$ & 4464.12& 4463& 4159 &$\ldots$  \\
     &$5{^1S_0}$&  4439.2 & $\ldots$ & 4440.12& 4446 &  $\ldots$ &  $\ldots$ \\
\hline

\end{tabular}
\end{center}
\end{table*}

\begin{table*}
\begin{center}
\tabcolsep 8pt
   \small
\caption{P-wave mass spectrum for $ b\bar{b}$  and $c\bar{c}$ bound states(in MeV).} \label{tab4}
\begin{tabular}{c c c c c c c c}
\hline
\hline
&&&Bottomonium \\
\hline

nL & State &  Present  &  Experimental \cite{pdg2017}  & \cite{prd2012} &\cite{radford}  & \cite{Wei-Jun Deng} & \cite{aeb}\\
\hline

1P&$1{^3P_2}$& 9912.3& 9912.21 $\pm$ 0.26   & 9907.89 & 9912.3 & 9921 &  9812   \\
  &$1{^3P_1}$& 9901.8 &  9892.78 $\pm$ 0.26  & 9887.63  & 9904.7 & 9903&  9812   \\
  &$1{^3P_0}$& 9889.2 &  9859.44 $\pm$ 0.42  & 9862.29  & 9861.39 & 9864& 9811     \\
  &$1{^1P_1}$& 9854.1 & 9899.3 $\pm$ 0.8             & 9896.07  & 9899.93 &9909&  9812 \\

2P&$2{^3P_2}$& 10265.9 & 10268.65 $\pm$ 0.22 & 10267.65 & 10271.2 & 10264 &  10044  \\
  &$2{^3P_1}$& 10258.9 & 10255.46 $\pm$ 0.22 & 10255.74 & 10254.8 & 10249& 10043      \\
  &$2{^3P_0}$& 10234.7 & 10232.50 $\pm$ 0.40 & 10240.85 & 10230.5 & 10220& 10042      \\
  &$2{^1P_1}$&  10264.9 & $\ldots$     & 10260.70 & 10261.8 & 10254&10043      \\

3P&$3{^3P_2}$ & 10516.9&   $\ldots$  & 10516.28 &$\ldots$ & 10528&  10272  \\
  &$3{^3P_1}$&10508.8  &    10512.1 $\pm$ 2.3  & 10507.24 & $\ldots$&10515& 10271    \\
  &$3{^3P_0}$&10497.6  &     $\ldots$ &  10497.07& $\ldots$ &10490&  10270   \\
  &$3{^1P_1}$& 10540.2  & $\ldots$ & 10511.30 & $\ldots$ & 10519& $\ldots$   \\

4P&$4{^3P_2}$& 10707.0 &  $\ldots$      & $\ldots$& $\ldots$ &   $\ldots$ &$\ldots$\\
  &$4{^3P_1}$& 10706.5 &    $\ldots$     & $\ldots$& $\ldots$&  $\ldots$ & $\ldots$\\
  &$4{^3P_0}$& 10703.8 &   $\ldots$     & $\ldots$&  $ \ldots$&  $\ldots$ & $\ldots$\\
  &$4{^1P_1}$& 10704.6  &  $\ldots$       &  $\ldots$&  $\ldots$&  $\ldots$ &  $\ldots$\\

\hline

&&&Charmonium\\
\hline

nL & State & Present  &  Experimental \cite{pdg2017} & \cite{prd2012}& \cite{Bai-Quing Li} & \cite{nr}& \cite{aeb}\\
\hline

1P &$1{^3P_2}$& 3554.2& 3556.20 $\pm$ 0.09  & 3570.00 & 3554 & 3556 &3467\\
   &$1{^3P_1}$& 3513.0 & 3510.66 $\pm$ 0.07  & 3490.94 & 3510 & 3505&3468\\
   &$1{^3P_0}$& 3418.4 & 3414.75 $\pm$ 0.31  & 3392.11 & 3433 & 3424&3468\\
   &$1{^1P_1}$& 3518.7& 3525.38 $\pm$ 0.11  & 3523.88 & 3519 & 3516&3467\\

2P &$2{^3P_2}$& 3921.2   &  3927.2 $\pm$ 2.6 & 3949.01 & 3937 & 3972&3815\\
   &$2{^3P_1}$& 3901.8   & $\ldots$ & 3902.55 & 3901 & 3925 &3815\\
   &$2{^3P_0}$& 3824.9   &  $\ldots$ & 3844.49 & 3842 &3852&3814\\
   &$2{^1P_1}$& 3956.2   & $\ldots$   & 3921.91 & 3908 &3934&3815\\

3P &$3{^3P_2}$& 4203.7& $\ldots$    & 4211.78& 4208  & 4317&4163\\
   &$3{^3P_1}$& 4174.6& $\ldots$   & 4178.47& 4178 & 4271&4162\\
   &$3{^3P_0}$& 4136.0 & $\ldots$    & 4136.84& 4131  &4202&4160\\
   &$3{^1P_1}$& 4231.1 & $\ldots$      & 4192.35& 4184  & 4279&$\ldots$ \\

4P &$4{^3P_2}$& 4415.1  & $\ldots$    & $\ldots$& $\ldots$  & $\ldots$ &$\ldots$ \\
   &$4{^3P_1}$& 4409.1  & $\ldots$   & $\ldots$ &  $\ldots$ &$\ldots$ &$\ldots$  \\
   &$4{^3P_0}$& 4383.2  &  $\ldots$  & $\ldots$& $\ldots$ &$\ldots$  &  $\ldots$ \\
   &$4{^1P_1}$& 4446.4  &  $\ldots$& $\ldots$ & $\ldots$  & $\ldots$ \\
\hline

\end{tabular}
\end{center}
\end{table*}

\begin{table*}
\begin{center}
\tabcolsep 8pt
   \small
\caption{D - wave mass spectrum for $ b\bar{b}$  and $c\bar{c}$ bound states(in MeV).} \label{tab5}
\begin{tabular}{c c c c c c c c}
\hline

&&&Bottomonium \\
\hline

nL & State &  Present  &  Experimental \cite{pdg2017}   & \cite{prd2012} & \cite{radford} & \cite{Wei-Jun Deng}& \cite{aeb}\\
\hline

1D&$1{^3D_3}$& 10140.4   & $\ldots$ &10176.68& 10163.1&  10157 &    9980    \\
  &$1{^3D_2}$& 10138.7   & 10163.7 $\pm$ 1.4 &10162.26& 10157.3& 10153 & 9980  \\
  &$1{^3D_1}$& 10136.0   &  $\ldots$  &10147.31& $\ldots$  &  10146 &  9980  \\
  &$1{^1D_2}$& 10068.2   & $\ldots$  &10166.00& 10158.6& 10153 &     9980   \\

2D &$2{^3D_3}$& 10398.7 & $\ldots$  &10447.09&  10455.7& 10436&   10175  \\
   &$2{^3D_2}$& 10397.1 & $\ldots$  &10437.52& 10450.3 & 10432&  10174   \\
   &$2{^3D_1}$& 10395.7 & $\ldots$  &10427.59&  $\ldots$& 10425&   10174 \\
   &$2{^1D_2}$& 10336.0 & $\ldots$  &10440   & 10451.4 & 10432&  10174  \\

3D &$3{^3D_3}$& 10620.9 &   $\ldots$ &10651.86&  $\ldots$ & $\ldots$&   $\ldots$  \\
   &$3{^3D_2}$& 10619.3 &   $\ldots$ &10644.62&  $\ldots$ &$\ldots$ &  $\ldots$   \\
   &$3{^3D_1}$& 10616.8&   $\ldots$  &10637.12&  $\ldots$ & $\ldots$ & $\ldots$   \\
   &$3{^1D_2}$& 10564.3 & $\ldots$   &10646.50& $\ldots$  & $\ldots$ &   $\ldots$ \\

4D &$4{^3D_3}$& 10820.9 &   $\ldots$  &10816.93& $\ldots$ & $\ldots$& $\ldots$   \\
   &$4{^3D_2}$& 10819.3 &   $\ldots$  &10811.09& $\ldots$ & $\ldots$&  $\ldots$  \\
   &$4{^3D_1}$& 10816.9 &    $\ldots$ &10805.03& $\ldots$ &$\ldots$&   $\ldots$  \\
   &$4{^1D_2}$& 10768.8 &  $\ldots$   &10812.60& $\ldots$ &$\ldots$&  $\ldots$ \\

5D &$5{^3D_3}$& 11005.2 &   $\ldots$  &10955.6 & $\ldots$ & $\ldots$&   \\
   &$5{^3D_2}$& 11003.7 &   $\ldots$  &10950.7 & $\ldots$ & $\ldots$&    \\
   &$5{^3D_1}$& 11001.4 &    $\ldots$ &10945.6 & $\ldots$ & $\ldots$ &     \\
   &$5{^1D_2}$& 10956.7 &  $\ldots$   &10952.0 & $\ldots$ & $\ldots$&     \\

\hline

&&&Charmonium\\
\hline

nL & State & Present  &  Experimental \cite{pdg2017} & \cite{prd2012}& \cite{Bai-Quing Li} & \cite{nr} & \cite{aeb}\\ \\
\hline

1D   &$1{^3D_3}$& 3769.6 & $\ldots$ &3843.95 & 3799   & 3806& 3805  \\
    &$1{^3D_2}$& 3756.1 &$\ldots$& 3787.72&  3798  &3800&  3807 \\
    &$1{^3D_1}$&  3745.3&  3773.13 $\pm$ 0.35 &3729.41&  3787  &3785&  3808  \\
     &$1{^1D_2}$&   3662.2&  $\ldots$ &3802.30& 3796    &3799& 3806   \\

2D   &$2{^3D_3}$&  4060.7 &$\ldots$ & 4132.53 & 4103  & 4167&  4143  \\
     &$2{^3D_2}$&  4048.4 &$\ldots$& 4095.17&  4100  & 4158&  4145   \\
    &$2{^3D_1}$&    4038.9 &$\ldots$ & 4056.43& 4089  & 4142& 4145  \\
     &$2{^1D_2}$&  3968.9& $\ldots$& 4104.86&  4099  & 4158&  4143   \\

3D  &$3{^3D_3}$ &   4317.2&$\ldots$& 4350.66& 4331    & $\ldots$&  $\ldots$ \\
     &$3{^3D_2}$&  4307.0&$\ldots$& 4322.44& 4327    & $\ldots$&  $\ldots$  \\
   &$3{^3D_1}$&   4300.6&$\ldots$& 4293.18&  4317    & $\ldots$&   $\ldots$  \\
    &$3{^1D_2}$&  4236.2&$\ldots$ & 4329.76&  4326  & $\ldots$ &  $\ldots$ \\

4D   &$4{^3D_3}$&  4552.0 & $\ldots$ & 4526.41 & $\ldots$ & $\ldots$&   $\ldots$ \\
     &$4{^3D_2}$&  4541.6 & $\ldots$ & 4503.63& $\ldots$  & $\ldots$ & $\ldots$  \\
    &$4{^3D_1}$ & 4533.6  & $\ldots$ & 4480.01& $\ldots$ &$\ldots$&  $\ldots$  \\
    &$4{^1D_2}$&  4477.6  & $\ldots$ & 4509.54& $\ldots$  & $\ldots$ &  $\ldots$  \\

5D   &$5{^3D_3}$&  4768.7 & $\ldots$ & 4673.96 & $\ldots$ & $\ldots$& $\ldots$   \\
     &$5{^3D_2}$&  4758.9 & $\ldots$ & 4654.80 & $\ldots$  & $\ldots$ &  $\ldots$ \\
    &$5{^3D_1}$ & 4751.6  & $\ldots$ & 4634.92& $\ldots$ &$\ldots$&   $\ldots$     \\
    &$5{^1D_2}$&  4699.9  & $\ldots$ & 4659.77& $\ldots$  & $\ldots$ & $\ldots$    \\
\hline

\end{tabular}
\end{center}
\end{table*}

\begin{equation}\label{eq:p}
M_{Q\bar{Q}} = E_D^Q + E_D^{\bar{Q}}-E_{cm}
\end{equation}

here, $E_{cm}$ in general can be state dependent which we absorb in our potential parameter $V_0$. Thus, making $V_0$ as state dependent.\\

In this calculations, we incorporate additionally, the j-j
coupling,
spin-orbit and tensor interactions of confined one gluon exchange potential (COGEP) \cite{pcv,epj,prd2014,prd2016}.
The mass of the state thus represented by $M_{^{2s+1}L_J}$ as \cite{epj,prd2014,prd2016},
 \begin{equation}\label{eq:q}
M_{^{2s+1}L_J} =M_{Q\bar{Q}}(n_1l_1j_1,n_2l_2j_2)+ \langle V^{j_1 j_2}_{{Q\bar{Q}}} \rangle+
\langle V^{LS}_{{Q\bar{Q}}} \rangle + \langle V^{T}_{{Q\bar{Q}}} \rangle
\end{equation}

where the j-j coupling term is expressed as  \cite{pcv,epj,amp1,epj,prd2014,prd2016},

 \begin{equation}\label{eq:r}
\langle V^{j_1 j_2}_{{Q\bar{Q}}} \rangle = \frac{\sigma\langle j_1 j_2 J M|\widehat{j_1}\widehat{j_2}| j_1 j_2 J M\rangle}
{(E_Q + m_Q)(E_{\bar{Q}} + m_{\bar{Q}})}
  \end{equation}

here, $\sigma$ is j - j coupling constant.\\
 $\langle j_1 j_2 J M|\widehat{j_1}\widehat{j_2}| j_1 j_2 J M\rangle $ contains the
  square of the Clebsch- Gordan coefficient. The spin orbit interaction and tensor
  interactions are expressed respectively as  \cite{pcv,epj,prd2014,prd2016,amp1},
\begin{eqnarray}\label{eq:s}
\langle V^{L S}_{{Q\bar{Q}}} \rangle &=& \frac{\alpha_s}{4} \frac{N_Q^2 N_{\bar{Q}}^2}
{(E_Q + m_Q)(E_{\bar{Q}} + m_{\bar{Q}})}
\frac{\lambda_Q\lambda_{\bar{Q}}}{2r} \\ &&
\otimes[\overrightarrow {r} \times (\widehat{p}_Q-\widehat{p}_{\bar{Q}})
.(\sigma_Q-\sigma_{\bar{Q}})] (D_0'(r)+2 D_1'(r)) \nonumber \\&&
+ [ [\overrightarrow {r} \times (\widehat{p}_Q + \widehat{p}_{\bar{Q}}).
(\sigma_i-\sigma_j)(D_0'(r) - D_1'(r)) \nonumber
\end{eqnarray}
 and
\begin{eqnarray}\label{eq:t}
\langle V^{T}_{{Q\bar{Q}}} \rangle &=& -\frac{\alpha_s}{4} \frac{N_Q^2 N_{\bar{Q}}^2}
{(E_Q + m_Q)(E_{\bar{Q}} + m_{\bar{Q}})}
\lambda_Q\lambda_{\bar{Q}} \\ &&
\otimes ((\frac{D_1''(r)}{3}-\frac{D_1'(r)}{3r}) S_{Q{\bar{Q}}}) \nonumber
\end{eqnarray}

where, $S_{Q \bar Q} = \left[ 3 (\sigma_Q. {\hat{r}})(\sigma_{\bar Q}.
{\hat{r}})- \sigma_Q . \sigma_{\bar Q}\right]$ and ${\hat{r}} = {\hat{r}}_Q -
{\hat{r}}_{\bar Q}$ is the unit vector in the relative coordinate  \cite{epj,prd2014,prd2016}.

 The running strong coupling constant $\alpha_s$ is computed as \cite{epj,prd2014,prd2016},
 \begin{equation}
 \alpha_s = \frac{4 \pi}{(11-\frac{2}{3}\  n_\emph{f})\log\left(\frac{E^2_Q}{\Lambda^2_{QCD}}\right)}
 \end{equation}
with $n_\emph{f}$ = 3 and $\Lambda_{QCD}$ = 0.250 GeV for charmonium and $n_\emph{f}$ = 4 and $\Lambda_{QCD}$ = 0.156 GeV for bottomonium. In Eqs. (20) the spin-orbit term has been split into symmetric $(\sigma_Q + \sigma_{\bar Q})$ and anti-symmetric $(\sigma_Q - \sigma_{\bar Q})$ terms.

We have adopted the  form of the confined gluon propagators which are given by \cite{pcv,amp1,epj,prd2014,prd2016};
\begin{equation}
D_0 (r) = \left( \frac{\alpha_1}{r}+\alpha_2 \right) \exp(-r^2 c_0^2/2)
\end{equation}
and
\begin{equation}
D_1 (r) =  \frac{\gamma}{r} \exp(-r^2 c_1^2/2)
\end{equation}

where $\alpha_1$ = 1.035, $\alpha_2$ = 0.3977, $c_0$ = 0.3418 GeV , $c_1$ = 0.4123 GeV,  $\gamma$ = 0.8639 are the fitted parameter as in  \cite{amp1}. Other model parameters employed in the present calculation are listed in Table \ref{tab2}.

The hyperfine splittings of ground and radial excitation of the bottomonium and charmonium are important for the study of the radiative transition amplitudes. The high precision experimental data have provided accurate description of the hyperfine and fine structure interactions of quarkonia. The hyperfine splitting for S-Wave and the ratio of spin orbit splitting  for P-Wave charmonium and bottomonium are given by equation (25) and (26) respectively.

\begin{equation}
\triangle M_{hf} (nS) = M(n^3S_1) - M(n^1S_0)
\end{equation}

\begin{equation}
R = \frac{M(^3P_2) - M(^3P_1)}{ M(^3P_1) -  M(^3P_0) }
\end{equation}

\begin{figure}
 \centering
\includegraphics[width=12cm]{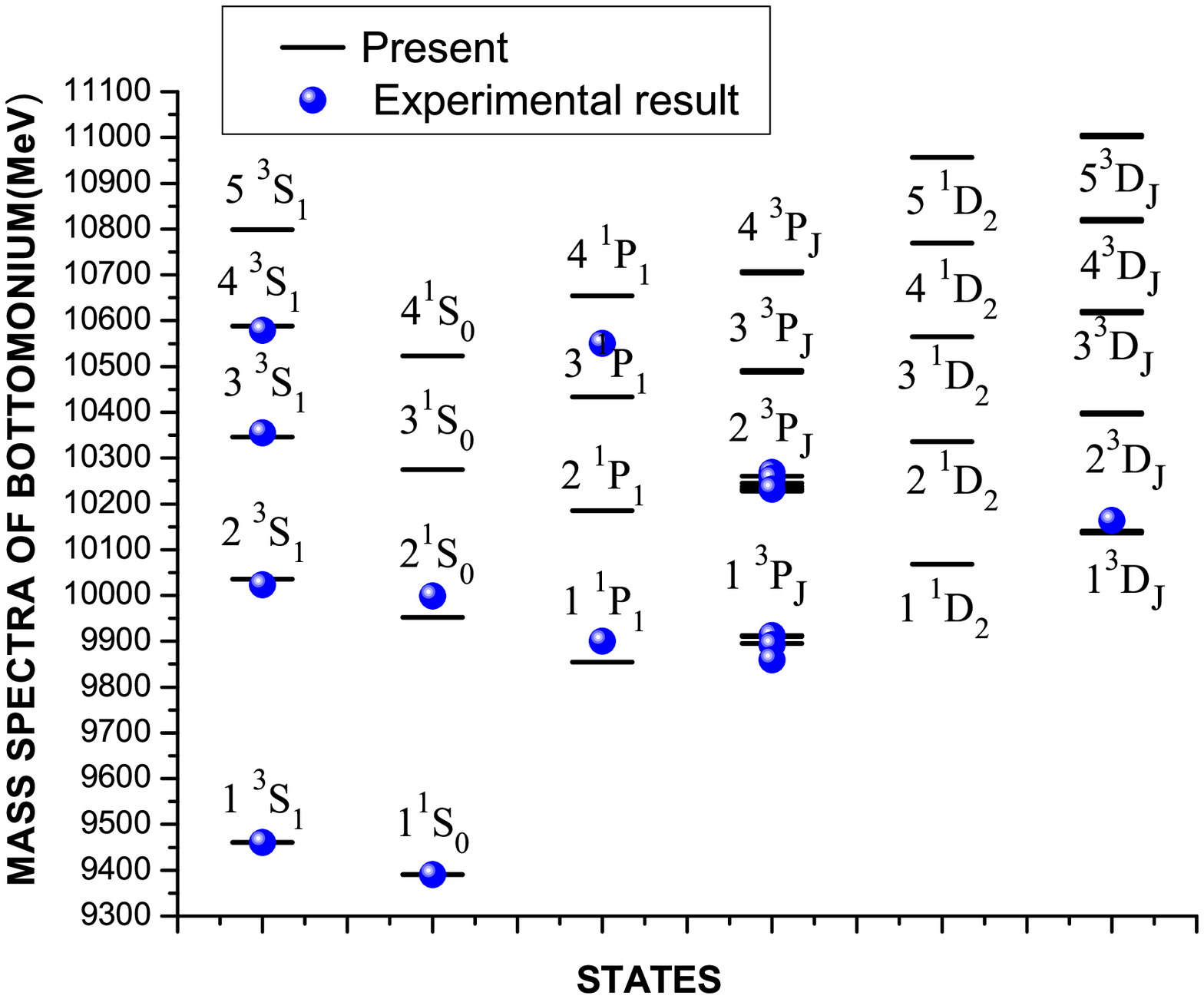}
\caption{\label{Mass spectrum of Bottomonium} Mass spectrum of Bottomonium}
\end{figure}

\begin{figure}
 \centering
\includegraphics[width=12cm]{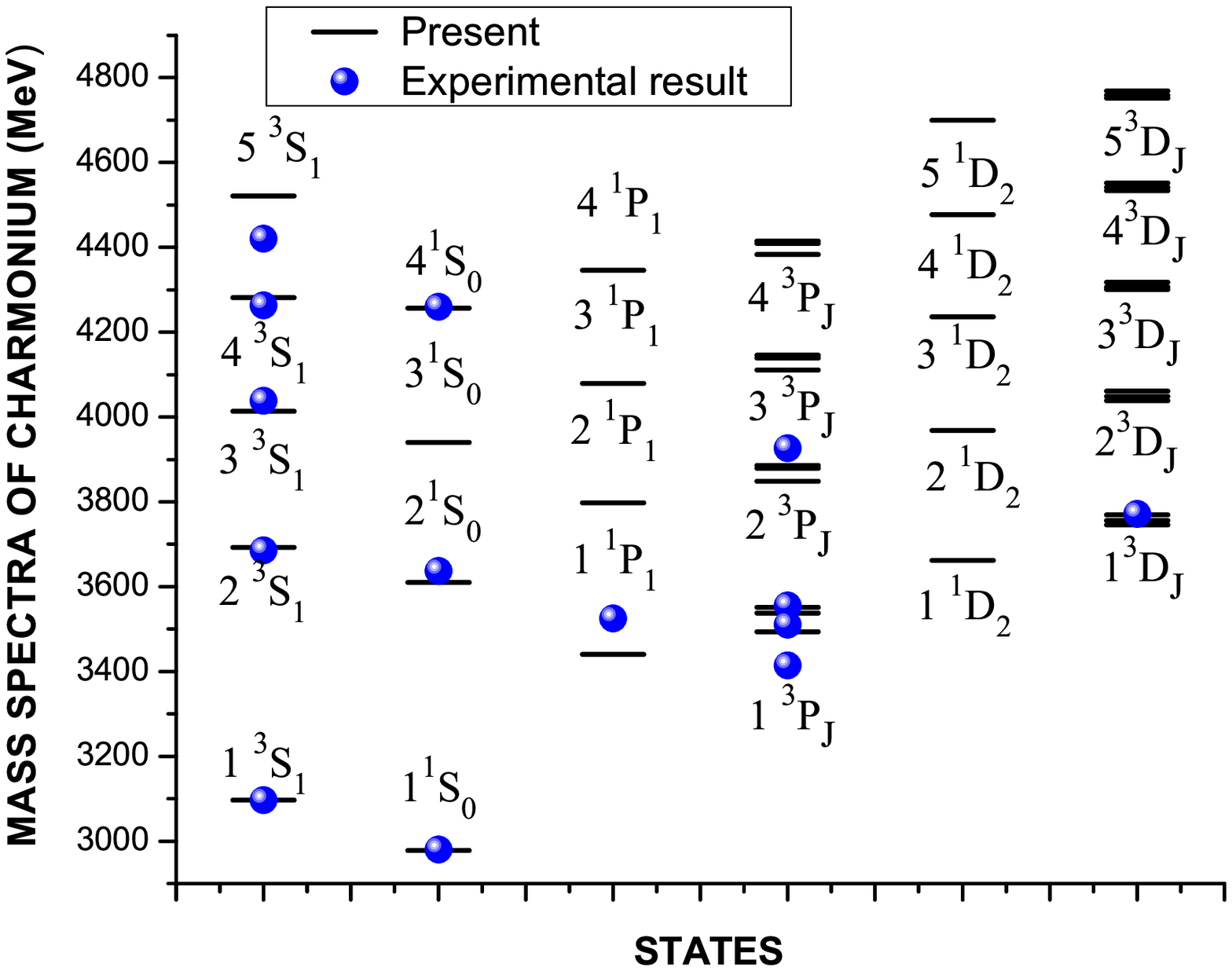}
\caption{\label{Mass spectrum of Charmonium} Mass spectrum of Charmonium}
\end{figure}
\paragraph*{} The computed S - wave, P - wave and D - wave mass spectra of bottomonium and charmonium are tabulated in Table \ref{tab3},  \ref{tab4} and \ref{tab5} . The corresponding energy level diagram are shown in Fig 1 and 2 respectively. The hyperfine splitting for S- wave and the spin orbit splitting ratio for P-wave is tabulated in Table \ref{tab11} and \ref{tab12}.

\begin{table*}
\begin{center}
\tabcolsep 5pt
   \small
\caption{The hyperfine splitting (in MeV) for S-wave bottomonium and charmonium.}\label{tab11}
\begin{tabular}{cccccc}
\hline

State (ns) & Hyperfine splitting & Present & Experimental \cite{pdg2017} & \cite{de}& \cite{prd2012}\\
\hline

&& bottomonium\\
\hline

1S & $\triangle M (1S) = M(1^3S_1) - M(1^1S_0)$ &   70.2      &  70 & 60& 68 \\

2S & $\triangle M (2S) = M(2^3S_1) - M(2^1S_0)$ &   24.8 $(\sim 25)$  & 24  &30&33\\

3S  & $\triangle M (3S) = M(3^3S_1) - M(3^1S_0)$ & 30.9 $(\sim 31)$ & $\ldots$ &27&22\\

4S  &  $\triangle M (4S) = M(4^3S_1) - M(4^1S_0)$ &  21.8 $(\sim 22)$      &  $\ldots$ &26&17  \\
5S  &  $\triangle M (5S) = M(5^3S_1) - M(5^1S_0)$ &  20.1     &  $\ldots$ & $\ldots$&  14\\
\hline

&& Charmonium\\
\hline

State (ns) & Hyperfine splitting & Present & Experimental \cite{pdg2017} & Lattice QCD \cite{tk} & NRp model \cite{tk}\\
\hline

1S &  $\triangle M (1S) = M(1^3S_1) - M(1^1S_0)$ & 118.9 $(\sim 119)$   &  116 &  114  & 108\\

2S & $\triangle M (2S) = M(2^3S_1) - M(2^1S_0)$  & 53.9 $(\sim 54)$  &   49 &  41  & 42\\

3S  &  $\triangle M (3S) = M(3^3S_1) - M(3^1S_0)$  & 31.6 &  $\ldots$&  25 &  29 \\

4S  &  $\triangle M (4S) = M(4^3S_1) - M(4^1S_0)$ &4.3 &  3 & $\ldots$  &$\ldots$\\
5S  &  $\triangle M (5S) = M(5^3S_1) - M(5^1S_0)$ & 2.3&  $\ldots$ &$\ldots$   &$\ldots$\\
\hline

\end{tabular}
\end{center}
\end{table*}

\begin{table*}
\begin{center}
\tabcolsep 5pt
\small
\caption{The ratios of spin orbit splitting (R) for P wave bottomonium and charmonium .}\label{tab12}
\begin{tabular}{ccccc}
\hline

State (nP) & Present & Experimental \cite{pdg2017}& \cite{de}& \cite{prd2012} \\
\hline

& & bottomonium\\
\hline

1P &  0.83   &  0.60 & 0.80& 0.80 \\

2P & 0.29   & 0.56  & 0.60& 0.80 \\

3P  &  0.72   &  $\ldots$ & 0.72& 0.78\\

4P  &  0.18   &  $\ldots$ &$\ldots$& $\ldots$ \\
\hline

& & Charmonium\\
\hline

State (nP) & Present & Experimental \cite{pdg2017} & Lattice QCD \cite{tk} & NRp model \cite{tk}\\
\hline

1P &  0.43   &  0.47 & 0.46 & 0.62 \\

2P &  0.25   &   $\ldots$ & $\ldots$& 0.64 \\

3P  &  0.75   &  $\ldots$ &$\ldots$& $\ldots$\\

4P  &  0. 23  &  $\ldots$ &$\ldots$& $\ldots$\\
\hline

\end{tabular}
\end{center}
\end{table*}

\begin{table*}
\begin{center}
\tabcolsep 5pt
   \small
\caption{Vector Decay Constant($F_v$ in MeV) of the S - wave and D - wave Bottomonium and  Charmonium states.} \label{tab6}
\begin{tabular}{c c c c c c c c}
\hline

&& Bottomonium\\
\hline

State &  Present  &  Experimental \cite{pdg2017}&\cite{Bhaghyesh} & \cite{Lakhina} &\cite{bhavin}&State & Present\\  \\
\hline

1S &   705.4 & 715 $\pm$ 5  &  831   &  665   & 867 & 1D & 208.3 \\
2S &  554.9  & 498 $\pm$ 8  &  566  &  475 &  673  & 2D& 181.3 \\
3S & 436.8   & 430 $\pm$ 4  &  507  &  418  & 595 & 3D& 151.4  \\
4S & 332.4  & 336 $\pm$ 18  &  481  &  388  & 549 &4D & 135.4\\
5S & 286.5  & $\cdots$      &    458   &   367    &    516 &5D & 113.1  \\
\hline

&& Charmonium \\
\hline

State &  Present  &  Experimental \cite{pdg2017}    & \cite{Bhaghyesh} & \cite{Lakhina} &\cite{rai}&State&Present\\
\hline

1S &   419.9 & 416 $\pm$ 6   &  462   &  393 & 589&1D& 102.5\\
2S &  285&  304 $\pm$ 4 &  369  &  293 &  328 &2D& 83.9\\
3S & 218 & $\cdots$  & 329   &  258 &   244 &3D&65.6\\
4S & 165.7 &  $\cdots$  & 310  & $\cdots$   & $\cdots$ &4D&54.2\\
5S & 106.2&  $\cdots$  & 290 & $\cdots$   & $\cdots$&5D&42.3 \\
\hline

\end{tabular}
\end{center}
\end{table*}

\begin{table*}
\begin{center}
\tabcolsep 5pt
   \small
\caption{Leptonic decay width (in keV) of the S - wave and D - Wave Bottomonium and Charmonium states.} \label{tab7}
\begin{tabular}{c c c c c c c c}
\hline

&&Bottomonium\\
\hline

State &  Present  &  Experimental\cite{pdg2017}  & \cite{prd2012}  & \cite{radford} & \cite{pcv} & state& Our \\
\hline

1S &  1.30 & 1.34 $\pm$0.018   & 1.203  &  1.33 & 1.809  &1D & 0.106\\
2S & 0.76  & 0.612 $\pm$ 0.011 & 0.519  &  0.62 & 0.797  &2D & 0.078\\
3S & 0.45  & 0.443 $\pm$ 0.008 & 0.330  &  0.48 & 0.618  &3D & 0.051\\
4S & 0.26  & 0.272 $\pm$ 0.029 & 0.241  &  0.40 &  0.541 &4D & 0.042\\
5S & 0.18  & $\cdots$          & 0.19  &  $\cdots$   &  $\cdots$ &5D & 0.028\\
\hline

&&Charmonium \\
\hline

State &  Present  &  Experimental\cite{pdg2017}  & \cite{prd2012}  & \cite{radford1} & \cite{pcv1} &State& Our\\
\hline

1S & 5.63 & 5.55 $\pm$ 0.14  &  4.94   &  1.89  & 5.469  &1D& 0.27 \\
2S & 2.19 & 2.48 $\pm$ 0.06  &  1.686  &  1.04   & 2.140 &2D& 0.17 \\
3S & 1.20 & $\cdots$         &  0.959  &  0.77  & 0.796  &3D&0.099\\
4S & 0.63 &  $\cdots$        &  0.654  &  0.65  & 0.288  &4D&0.064\\
5S & 0.24 &  $\cdots$        &  0.489  &   $\cdots$   &  $\cdots$  &5D&0.044\\
\hline

\end{tabular}
\end{center}
\end{table*}

\begin{table*}
\begin{center}
\tabcolsep 2.1pt
\small
\caption{Mixing angle and the leptonic decay widths of S - D wave admixture states.}\label{tab9}
\begin{tabular}{cccccccccc}
\hline

Experimental 	&	$J^P$	&&	mixed state 	&	 \% mixing  	&	\multicolumn{2}{c}{Mass of mixed state (MeV)}	&		&	\multicolumn{2}{c}{Mixed state leptonic }			\\
	state&		&&	configuration	&of S-wave&&		&&&	decay width (keV)					\\
\cline{6-7}
\cline{9-10}
	&		&&		&		&		&		&		&		\\
	&		&&		&		&	Our 	&	Experimental	&&	Our 	&	Experimental	\\
\hline
& && && Charmonium like states\\
\hline
Y(4008)	&	$1^-$	&&	${2^3S_1}$ and ${2^3D_1}$ 	&	8.6$\%$ 	&	4008.4	&	$4008^{+121}_{-49}$ \cite{cz}	&&	0.347	&	0.862 $\pm$ 0.241 \cite{a1}	\\

$\psi(4160)$	&	$1^-$	&&	${3^3S_1}$ and ${3^3D_1}$	&	39.2$\%$ 	&	4191	&	$4191 \pm 5$ \cite{pdg2017}	&&	0.534	&	$0.48 \pm  0.22$ keV \cite{pdg2017}	\\

Y(4220)	&	$1^-$	&&	${3^3S_1}$ and ${3^3D_1}$	&	28.7$\%$ 	&	4220.7	&	4222.0$\pm$ 3.1$\pm1.4$  \cite{bes}	&&	0.417	&	NA	\\

X(4260)	&	$1^-$	&&	${3^3S_1}$ and ${3^3D_1}$	&	14.41$\%$ 	&	4260.5	&	$4251 \pm 9 $ \cite{pdg2017}	&&	0.258	&	NA	\\

X(4360)	&	$1^-$	&&	${4^3S_1}$ and ${4^3D_1}$ 	&	64.97$\%$ 	&	4360	&	$4346 \pm 6$ \cite{pdg2017}	&&	0.431	&	$<$ 0.57 eV	\cite{pdg2017}\\

X(4630)	&	$1^-$	&&	${5^3S_1}$ and ${5^3D_1}$ 	&	37.92$\%$ 	&	4634	&	$4634^{+9}_{-11}$ \cite{gp1}	&&	0.117	&	NA	\\

X(4660)	&	$1^-$	&&	${5^3S_1}$ and ${5^3D_1}$ 	&	34.37$\%$ 	&	4645	&	4643$\pm$ 9 \cite{pdg2017}	&&	0.110	&	$<$0.45 eV \cite{pdg2017}	\\
	
\hline
& && && Bottomonium like state\\
\hline
$\Upsilon(10860)$	&	$1^-$	&&	${5^3S_1}$ and ${5^3D_1}$ 	&	45.45$\%$ 	&	10880.1	&	$10891  \pm  4$   \cite{pdg2017}	&&	0.096	&	$0.31 \pm  0.07$ keV \cite{pdg2017}	\\
\hline
*NA = Not Available
\end{tabular}
\end{center}
\end{table*}

\begin{table*}
\begin{center}
\tabcolsep 4.0pt
\small
\caption{masses of mixed P-wave +ve parity states.} \label{tab8}
\begin{tabular}{ccc}
\hline

Experimental state & Mixed state configuration & Present(MeV) \\
\hline

& Charmonium like states\\
\hline

X(3940)&${2^3P_1}$ and ${2^1P_1}$& 3939.06\\
X(4020)&${2^3P_1}$ and ${3^1P_1}$& 4011.56\\
X(4140)&${2^3P_1}$ and ${3^1P_1}$& 4121.33\\
X(4350)&${4^3P_1}$ and ${3^1P_1}$& 4349.76\\
\hline

& Bottomonium like state\\

\hline
X(10610)&${4^3P_1}$ and ${3^1P_1}$& 10595.63\\
\hline
\end{tabular}
\end{center}
\end{table*}

\begin{table*}
\begin{center}
\tabcolsep 5pt
   \small
\caption{The ratios of  $\frac{\Gamma_{ee}(nS)}{\Gamma_{ee}(1S
 )}$ for bottomonium and charmonium states.}\label{tab10}
\begin{tabular}{cccc}
\hline

$\frac{\Gamma_{e^+e^-}(\Upsilon(nS)}{\Gamma_{e^+e^-} (\Upsilon(1S)}$ & Present & Experimental \cite{pdg2017} & \cite{hff}\\
\hline

& bottomonium\\
\hline

$\frac{\Gamma_{e^+e^-}(2S)}{\Gamma_{e^+e^-}(1S)}$  &  0.58   &  0.46 & 0.50  \\
\hline
$\frac{\Gamma_{e^+e^-}(3S)}{\Gamma_{e^+e^-} (1S)}$  &  0.35   & 0.33  & 0.36   \\
\hline
$\frac{\Gamma_{e^+e^-}(4S)}{\Gamma_{e^+e^-} (1S)}$  &  0.20   &  0.20 &  0.29  \\
\hline
$\frac{\Gamma_{e^+e^-}(5S)}{\Gamma_{e^+e^-} (1S)}$  &  0.13 &  $\cdots$ &  0.24  \\
\hline
\hline
& Charmonium\\

\hline
$\frac{\Gamma_{e^+e^-} (\psi(nS)}{\Gamma_{e^+e^-} (\psi(1S)}$ & Present &  Experimental \cite{pdg2017}&  \cite{hff}\\
\hline

$\frac{\Gamma_{e^+e^-}(2S)}{\Gamma_{e^+e^-}(1S)}$  &  0.39   &   0.43   & 0.48  \\
\hline
$\frac{\Gamma_{e^+e^-}(3S)}{\Gamma_{e^+e^-}(1S)}$  &  0.21   & $\cdots$    & 0.32  \\
\hline
$\frac{\Gamma_{e^+e^-}(4S)}{\Gamma_{e^+e^-}(1S)}$  &  0.11  &   $\cdots$   & 0.24  \\
\hline
$\frac{\Gamma_{e^+e^-}(5S)}{\Gamma_{e^+e^-}(1S)}$  &  0.04 & $\cdots$   &0.19  \\
\hline

\end{tabular}
\end{center}
\end{table*}

\section{Decay constants and Leptonic Decay width of $1^{--}$ quarkonia}
\paragraph*{} The leptonic decay width is a tool to understand the compactness
of the mesonic system. We know that leptonic decay width of $J/\Psi$ is reasonably predicted by the phenomenological model. At the same time heavy quarkonium states are
precisely most sensitive to the short range one gluon exchange interaction between quarks and antiquarks \cite{Louise}.
\paragraph*{} In relativistic quark model, the vector decay constant is expressed through meson wave function $f(\overrightarrow{q})$ in  momentum space as given by \cite{ktc};\\

\begin{equation}
f_V  = \frac{2\sqrt{3}}{M}\int d^3q \left (\frac{ m + E }{E}-\frac{\overrightarrow{q}^2}{3E^2}) \right)f(\overrightarrow{q})
\end{equation}

Where, $E = \sqrt{\overrightarrow{q}^2 + m^2}$ and $\sqrt{3}$ is the color factor. M is the mass of vector state. The leptonic decay width is expressed as \cite{ktc},

\begin{equation}
\Gamma(V\rightarrow e^+ e^-) =\frac{4}{3} \pi \alpha^2 e_Q^2 \frac{f_V^2}{M}
\end{equation}

\paragraph*{} The computed decay constants and leptonic decay widths in the case of charmonium and bottomonium sates are presented in Table \ref{tab6} and \ref{tab7} respectively. The ratio of $\frac{\Gamma_{ee}(nS)}{\Gamma_{ee}(1S
 )}$ for bottomonium and charmonium states are listed in Table \ref{tab10}. Along with the mass predictions, the leptonic decay widths are also important for the identification of the structures of quarkonia-like states.

\section{Quarkonia - like states as mixed quarkonia states} \label{intro}

\paragraph*{}It is known that many of the hadronic states which are observed and yet not clear about their structure can be the admixture of the nearby iso-parity states. In general, the mass of a mixed state ($M_{nL}$) can be expressed in terms of the two mixing states ($nl$  and $n'l'$) as
\begin{equation}\label{202}
M_{nL} = \mid a^2\mid M_{nl} + ( 1-\mid a^2\mid ) M_{n'l'}
\end{equation}
 Where,$\mid a^2\mid$  $=$ $\cos^2 \theta$ and $\theta$ is mixing angle. With the help of this equation we can obtain mixed state configuration and mixing angle \cite{prd2012}.\\
The computed masses and their leptonic decay width of the S-D wave admixture states are presented in Table \ref{tab9}.\\
In this context we consider the admixture of nearby P-waves for the predictions of some of the $1^+$ states and for other $1^-$ states we consider the S-D wave mixing
 \cite{epj,epjC,epl}. The mixed P wave states can be expressed as \cite{epj,epjC,epl},
\begin{equation}
|\alpha\rangle=\sqrt{\frac{2}{3}}|^3P_1\rangle+\sqrt{\frac{1}{3}}|^1P_1\rangle
\end{equation}
\begin{equation}
|\beta\rangle=-\sqrt{\frac{1}{3}}|^3P_1\rangle+\sqrt{\frac{2}{3}}|^1P_1\rangle
\end{equation}
Where, $|\alpha\rangle$, $|\beta\rangle$ are states having same parity.
We can write the masses of these states in terms of
the predicted masses of the pure P wave states ($^3P_1$ and $^1P_1$) as \cite{epj,epjC,epl}:

\begin{equation}
M(|\alpha\rangle)=\frac{2}{3}M(^3P_1)+\frac{1}{3}M(^1P_1)
\end{equation}

\begin{equation}
M(|\beta\rangle)=\frac{2}{3}M(^1P_1)+\frac{1}{3}M(^3P_1)
\end{equation}

\paragraph*{} The computed mixed P - wave states for the positive parity quarkonia - like states are listed in Table \ref{tab8} and the experimental states which are close to these mixed states are also listed for comparison.


\section{Result and discussion}

\paragraph*{} In the framework of the Dirac relativistic  quark model,
we  have studied the mass spectrum of bottomonium - like and Charmonium -like states. To obtain these mass spectra we have solved Dirac equations with a linear plus constant confinement potential. In our calculations, spin-dependent interactions are included to remove degeneracy of the states. The predicted bottomonium  and charmonium spectra are in good agreement with the experimental data and
other available theoretical data. We have also predicted the 4S and 5S states for charmonium and bottomonium and
compared them with the available theoretical results. The predicted masses of the S-wave bottomonium states $3^3S_1$ (10356.2 MeV) and $4^3S_1$ (10576.2 MeV) are in accordance
with experimental results as quoted in the particle data group (PDG 2016) \cite{pdg2017}. The predicted masses of the
S-wave charmonium states $3^1S_0$ (3990.8 MeV), $4^1S_0$ (4262.1 MeV) and $5^1S_0$
(4439.2 MeV) are in accordance with other model
predictions \cite{prd2012,Bai-Quing Li,nr}.  The
computed P-wave bottomonium states $2^3P_2$ (10265.9 MeV),
$2^3P_1$ (10258.9 MeV) and $2^3P_0$ (10234.7 MeV) are in good
agreement with experimental \cite{pdg2017} results of $10268.65 \pm 0.22$ MeV,
$10255.46 \pm 0.22$ MeV and $10232.50 \pm 0.40 $ MeV respectively. For charmonium, the predicted P-wave mass of 3921.2 MeV for $2^3P_2$ state is in very good agreement with the  available experimental result of 3927.2 $\pm$ 2.6 MeV \cite{pdg2017}. The masses of 3P, 4P and 1D to 4D states and their fine structure splitting for bottomonium and charmonium  are fairly in good agreement with the available experimental
results and  other theoretical predictions. The predicted states such as $\chi_b(3P)$, $\chi_b(4P)$, $\Upsilon(3D)$, $\Upsilon(4D)$ and $\Upsilon(5D)$ of bottomonium and $\chi_c(2P)$ to $\chi_c(4P)$, 2D to 5D states of charmonium are not available experimentally.\\
\paragraph*{}In this paper, we have also examined the vector decay constant and leptonic decay widths for nS states of Bottomonium and charmonium within the relativistic framework. Decay widths of $\Upsilon (nS)\rightarrow e^+e^-$  and $\psi(nS)\rightarrow e^+e^-$ are shown in Table \ref{tab7}.  All the results for vector decay constant and leptonic decay width are calculated without QCD corrections. The calculated leptonic
decay widths for $\Upsilon(3S)$ and $\Upsilon(4S)$ are in good agreement with the available
experimental data, but the calculated values for $\Upsilon(2S)$
is slightly higher than the experimental value. In the case of $\psi(3S)$ state of charmonium, vector decay constant and leptonic decay width are slightly higher than the experimental result. Our results for $f_{J/\psi(1S)}$ and $f_{\psi(2S)}$ are in good agreement with the lattice QCD results of $399 \pm 4$ GeV and $143 \pm 81$ MeV respectively \cite{jjd}. It is observed that the leptonic decay widths decrease with radial excitations. From this we can conclude that the relativistic treatment is important for higher excited states.\\

\paragraph*{} To understand the structure of some of the newly found `X Y Z' quarkonia like - states with hidden charm and hidden bottom flavors, we consider here the possibilities for the mixing of $^3P_1$ and $^1P_1$ and of $^3S_1$ and $^3D_1$ iso-parity states. The calculated mixed states of ${^3P_1}$ - ${^1P_1}$ and ${^3S_1}$ - ${^3D_1}$ are listed in Table \ref{tab9} and \ref{tab8} respectively. The corresponding leptonic decay widths of the $1^{--}$ admixture states are also listed in Table \ref{tab9}.\\

\paragraph*{}Now, we briefly summarize the structure of some of the newly found quarkonia - like states based on our results of the masses and the leptonic decay widths.\\

$\bullet$ The X(3940)  has been observed in the $D^\ast\bar{D}$ channel with $J^P$ value $1^+$ but not in the $D\bar{D}$ decay mode \cite{ws}. Looking in to its parity, the possible identification of this state could be one of the  charmonium-like states. However the predicted 2P states are lower than this mass while the 3P states are slightly higher than this mass. Based on our analysis of mixed states, we predict X(3940) as $1^+$ state with admixture of $2^3P_1$ and $2^1P_1$ charmonia .\\

$\bullet$ Similarly, we predict X(4020) as an admixture of  $2^3P_1$ and $3^1P_1$ state with $J^P$ = $1^+$ having mass 4011.56 MeV.\\

$\bullet$ The Particle Data Group has renamed Y(4140) to X(4140) \cite{pdg2017}. Many attempts are made to study this state. According to ref \cite{zwg} this state might be a candidate of tetra quark state but because of unknown value of $J^P$,
its status is still not confirmed. According to our analysis it does not fit in to the admixture of P-states. But it fits well with the pure charmonium $3^3P_0$ state where by predicting its $J^P$ value as $0^+$.\\

$\bullet$ The X(4350) state whose $J^P$ value is not known experimentally is predicted as an admixture of $4^3P_1$ and $3^1P_1$ states having mass 4349.76 MeV or it might be pure charmonium $4^1P_1$ state with $J^P$ value $1^+$. Similarly, the X(10610) is also found to be the admixture of $4^3P_1$ and $3^1P_1$ states of bottomonium having mass 10595.63 MeV with $J^P$ equal to $1^+$. \\

$\bullet$ For the better identification of some of the $1^{--}$ states, we have considered both the predicted masses and their leptonic decay widths
in accordance with our previous study \cite{prd2012} as the successive mass differences [(n-1)S - nS] as well as the leptonic decay widths of $1^{--}$ states of quarkonia have shown to follow a specific decreasing pattern, characteristic of the bound state. \\

$\bullet$ Accordingly, our predicted mass for $\psi(3S)$ state as 4022.4 MeV and its leptonic decay width as 1.21 keV do not follow with the experimental mass of $\psi(3770)$ state and its experimental leptonic decay width of $0.262 \pm 0.018$ keV \cite{pdg2017}. But according to our predicted mass of $1^3D_1$  state as 3745.3 MeV and its computed leptonic decay width of 0.27 keV  indicate that $\psi(3770)$ is right candidate for a pure charmonium $1^3D_1$ state.\\

$\bullet$ In 2013,  Belle Collaboration updated the analysis of $ e^+ e^- \rightarrow J/\psi \pi^+ \pi^- $ with a 967 $fb^{-1}$ data sample and shows the invariant mass distribution of $J/\psi \pi^+ \pi^- $, the distribution is fitted with two coherent resonances. The existence of Y(4008) state was further confirmed \cite{b11}  in consistent with Belle's previous results \cite{cz}. There are some theoretical approaches to understand the structure of Y(4008) after Belle's confirmation \cite{cz}. X. Liu discussed some possibilities for the Y(4008), including both the (3S) charmonium states and the $D^\ast\bar{D}^\ast$ molecular state. For both possibilities, he found the branching ratio of $Y(4008) \rightarrow J/\psi \pi^0 \pi^0 $  is comparable with that of $Y(4008) \rightarrow J/\psi \pi^+ \pi^- $ \cite{a1}. Li and Chao studied the higher charmonium states in the non-relativistic screened potential model, and they predicted the Y(4008) as the (3S) charmonium state \cite{b1}. Chen, Ye, and Zhang found that the Y(4008) is difficult to identify $3^3S_1$ pure charmonium \cite{c1}. The Y(4008) was studied by Maiani, Piccinini, Polosa, and Riquer in their “type-II” diquark-antidiquark model, and interpreted as tetraquark state \cite{d1} and they have also assigned Y(4360) as  the first radial excitations of Y(4008) tetraquark state. In \cite{e1}, Zhou, Deng, and Ping also predicted the Y(4008) as a tetraquark state $cq\bar{c}\bar{q}$ with  $J^{PC} = 1^{- -}$ and $n^{2S+1}L_J$ = $1^1P_1$. They have used a color flux-tube model with a
four-body confinement potential to interpret the status of Y(4008). According to Dian-Yong Chen, the Fano-like interference induces an extra broad structure in $Y(4008) \rightarrow  \pi^+ \pi^- J/\psi$ as a companion peak to Y(4260) and also it explained why  Y(4008), Y(4260) and Y(4360) are absent in the experimental data of the R value scan \cite{f1} and they have  concluded that appearance of  Y(4008) peak is due to the interference of $\psi(4160)/\psi(4415)$ with the continuum of  $Y(4008) \rightarrow  \pi^+ \pi^- J/\psi$ \cite{f1}. However, very recently BESIII could not confirm the existence of Y(4008) \cite{bes}. In this context, Y(4008) is still a controversial state.\\

In the present study, the predicted mass of Y(4008) is found to be close to $2^3D_1$ charmonium state with just 8.6 $\%$ mixing with the $2^3S_1$ charmonium state. However, the computed leptonic decay width of the admixture state (0.347 keV) is much lower than the experimentally reported value of 0.86 keV \cite{a1}. Thus by considering both the mass and the leptonic decay width together, it is difficult to confirm or understand the structure of Y(4008). We look forward more refined experimental data for better understanding of this state.\\

$\bullet$ According to the present study, the state $\psi(4160)$ is found to be as an admixture of $3^3D_1$(60.8 $\%$) and $3^3S_1$ (39.2 $\%$) states with its leptonic decay width as 0.534 keV which is in accordance  with the experimental result of $0.48\pm0.22$ keV \cite{pdg2017}.\\

$\bullet$ The state Y (4260) was observed by the BaBar Collaboration in the $ J/\psi \pi^+ \pi^- $ channel in the initial state radiation (ISR) process \cite{ba1}. It was confirmed by CLEOc \cite{qne}, Belle \cite{cz} and an additional analysis done by BaBar \cite{lee}, with mass values varying in different analyses. The decay modes of the Y (4260) into $J/\psi$ and other charmonium states indicate the presence of a $c\bar{c}$ content. From PDG \cite{pdg2017}, the masses of some radial excitations, $\psi(2S)$ and $\psi(1P)$ are well established but masses of $\psi(3S)$, $\psi(4S)$, $\psi(1P)$, $\psi(2P)$ and $\psi(1D)$ still need more experimental investigation. Some theoretical interpretations for the Y (4260) are: hybrid mesons (mixing of $c\bar{c}$ and $c\bar{c}g$) \cite{p11,q11,r11}, tetraquark state \cite{s11}, hydrocharmonium \cite{t11,u11}, hadronic molecules of $\bar{D}D_1(2420) + c.c$ \cite{v11}, $\omega\chi_{c0}$ \cite{w11} etc. For the $\omega\chi_{c0}$ molecule, the predicted leptonic decay width is only about 23 eV \cite{w11}.  According to Felipe J. Llanes-Estrad, Y(4260) was proposed to be a conventional charmonium $\psi(4S)$ state and also estimated its leptonic decay width as 0.2 - 0.35 keV \cite{x11}.  Wen Qin, Si-Run Xue and  Qiang Zhao have predicted the upper limit of the Y (4260) leptonic decay width about 500 eV \cite{y11}. The LQCD also predicts the leptonic decay width as $ < 40$ eV for a hybrid charmonium state \cite{z11}.\\

According to new results from BESIII \cite{bes}, Y(4260) is not a simple peak. This measurement of the $e^+ e^- \rightarrow \pi^+ \pi^- J/\psi$ cross section was done by using both a small number of high-statistical data points and a large  number of low- statistics data points \cite{bes}.  They found resonance Y(4260) is described as combination of two peaks Y(4220) and Y(4330) \cite{bes}. However, the structure and interpretations of Y(4220) and Y(4330) are not yet understood.
Recently, X. Y. Gao, C. P. Shen and C. Z. Yuan  have predicted the value of leptonic decay width for Y(4220) can be as large as 200 eV or even higher based on current information \cite{xyg}. So the peaks observed by BESIII will provide more information about their structure. Thus the states, Y(4260), Y(4220) and Y(4330) have opened up new challenges in the charm sector.

\paragraph*{} According to latest PDG 2016 \cite{pdg2017}, the earlier state Y(4260) is now renamed as X(4260). We have analysed the status of X(4260), Y(4220) and Y(4330) states. \\

According to the present study Y(4220) state do not fit to be a pure charmonium state but fit to be an admixture of ($3^3D_1$)(71.3 $\%$) and ($3^3S_1$)(28.7 $\%$) states with its estimated leptonic decay width as 0.417 keV.\\

The second resonance reported by BES III, Y(4330) with mass $4326.8 \pm 10$ MeV is close to our predicted $3^3D_1$ state having mass 4300.6 MeV and its predicted leptonic decay width as 0.099 keV.\\

If we now consider X(4260) as pure $\psi(4S)$  state with the predicted mass equal to 4266.4 MeV then its leptonic decay width is predicted as 0.63 keV which is higher than the upper limit of 0.500 keV \cite{y11}. And if we consider X(4260) as the mixed states of Y(4220) and Y(4330) with a mixing  probability of $0.67 : 0.33$, then its leptonic decay width has to be 0.258 keV. The recent experimental measurements of BESIII \cite{u1,v1} suggests comparatively very small leptonic decay width for X(4260)\cite{w11}. Thus, X(4260) state can neither be identified as a pure state nor a mixed state. X(4260) state might be exotic state or hadronic molecular state. We require more experimental data for the confirmation of X(4260) state.\\

$\bullet$ Another controversial state is Y(4360) having $J^{P}$ value ${1^{-}}$ has now been renamed as X(4360) \cite{pdg2017}. This state might be a diquark-antidiquark type tetraquark state or it may be a mixed S-D wave charmonium states \cite{zwg1}. Present analysis suggests  it to be a mixed $4^3S_1$ and $4^3D_1$ charmonium state with leptonic decay width of 0.431 keV.\\

$\bullet$ X(4630) is compatible with our $1^{--}$ mixed charmonium
 state with admixture of $5^3S_1$ and$5^3D_1$ states having mass 4634 MeV and its predicted leptonic decay width as 0.117 keV.\\

$\bullet$  Structure of Y(4660) is also interesting because this state  was neither observed in $e^+e^-$ $\rightarrow$ $\gamma_{ISR} p^+ p^- J/\psi$ process, nor in the
mass distributions of a $c\bar{c}$ in the final stage of $e^-e^+$
collision experiment \cite{fkg}. Other theoretical approaches suggested it to be a molecular like structure.
According to latest PDG \cite{pdg2017}, Y(4660) has been renamed as X(4660)and its  $J^P $ = $1^-$ value suggests that it might be an admixture of $5^3S_1$ and $5^3D_1$ state  and we have predicted its leptonic  decay width to be 0.110 keV. However, the status of  these states is still a mystery and to resolve this we need more experimental results on their leptonic decay widths.\\

$\bullet$ The Particle Data Group has renamed $Y_b(10888)$ to $\Upsilon(10860)$ \cite{pdg2017}. In our present study by considering both its mass and leptonic decay width, we find it very difficult to assign it as the $b\bar{b}$, 5S state. Even if we consider it as an admixture of $5^3S_1$ and $5^3D_1$ state, its leptonic decay width estimated to be equal to 0.096 keV which is much lower than the experimentally reported value of 0.31 $\pm$ 0.07 keV \cite{pdg2017}. So, the status of  $Y_b(10888)$ or $\Upsilon(10860)$ as a conventional bottomonium state or an admixture of S-D states is doubted. More refined experimental observations of $\Upsilon(10860)$ can throw more light towards the understanding of this state.

\section{Summary}

In the present paper we have proposed a quark
model for hadrons. The approach is attractive due to its simplicity in applications to  quarkonia and exotic hadrons. In our model for meson mass spectroscopy we have solved Dirac equation to obtain the binding energy for individual quark/antiquark. Further the masses of the bound state is computed by adding the binding energies of the quark and antiquark with the addition of a center of mass correction.

In the last few years many states have been observed at B-factories (BaBar, Belle and CLEO), at proton-proton colliders (ATLAS, CMS, CDF, D0,
LHCb) and also at $\tau$ -charm facilities (CLEO-c, BES3) in the heavy quarkonium sector.
These charmonium-like and bottomonium-like states have provided new challenges for theorists as well as for experimentalists because it reveals the inner mechanisms of hadrons. There is no confirmation regarding XYZ states as an exotic states, molecular states and hybrid structure. We have predicted the status of few unknown states as an admixture of two states having same $J^P$ values and predicted their leptonic decay widths. The  LHCb (CERN), BES-III (China), PANDA (FAIR, Germany; after 2018) and Belle (Japan) experiments are expected pour more data in the quarkonium sector. With the help of advance experimental facilities we hope to get valuable information related to the newly observed hadronic states. The theoretical predictions will be helpful for experimental exploration of the hadronic states in the quarkonium sector.\\

\section*{Acknowledgments}
We acknowledge the financial support from DST-SERB, India (research Project number: SERB/F/8749/2015-16)




\end{document}